\documentclass[twocolumn,showpacs,preprintnumbers,amsmath,amssymb,prl,superscriptaddress]{revtex4}

\usepackage{epsfig}
\usepackage{graphicx}
\usepackage{dcolumn}
\usepackage{bm}

\begin{document}

\title{Classification of the electronic correlation strength in the Fe-pnictides: \\
The case of the parent compound BaFe$_{2}$As$_{2}$}

\author{S.L.~Skornyakov}
\author{A.V.~Efremov}
\author{N.A.~Skorikov}
\author{M.A.~Korotin}
\author{Yu.A.~Izyumov}
\author{V.I.~Anisimov}
\affiliation{Institute of Metal Physics, Russian Academy of Sciences,
620041 Yekaterinburg GSP-170, Russia}
\author{A.V.~Kozhevnikov}
\affiliation{Joint Institute for Computational Sciences, Oak Ridge
National Laboratory, P.O. Box 2008, Oak Ridge, Tennessee 37831-6173,
USA}
\author{D. Vollhardt}
\affiliation{Theoretical Physics III, Center for Electronic
Correlations and Magnetism, Institute for Physics, University of
Augsburg, Augsburg 86135, Germany}

\date{\today}

\begin{abstract}
Electronic correlations in the Fe-pnictide BaFe$_{2}$As$_{2}$ are
explored within LDA+DMFT, the combination of density functional
theory with dynamical mean-field theory. While the correlated band
structure is substantially renormalized there is only little
transfer of spectral weight. The computed $\bf k$-integrated and
$\bf k$-resolved spectral functions are in good agreement with
photoemission spectroscopy (PES) and angular resolved PES
experiments. Making use of a general classification scheme for the
strength of electronic correlations we conclude that
BaFe$_{2}$As$_{2}$ is a moderately correlated system.

\end{abstract}


\maketitle

The recent discovery of an entirely new class of high-$T_c$
superconductors based on quasi-twodimensional FeAs-layers~\cite
{Kamihara-08} rather than on CuO-layers has stimulated intense
experimental and theoretical activity. In  striking similarity with
the high-$T_c$ cuprates, the undoped pnictides are not
superconducting under ambient pressure, but exhibit an
antiferromagnetic commensurate spin-density wave below 150~K~\cite
{neutrons}. When electrons (or holes) are added to the system via
doping antiferromagnetism is suppressed and superconductivity
appears. Since it is generally accepted that the strong Coulomb
correlations between the Cu $3d$ electrons are responsible for the
anomalous properties of cuprates, it is tempting to suggest that the
same is true for the Fe $3d$ electrons in pnictides.

Investigations of the Coulomb correlation strength in the
Fe-pnictides have arrived at different conclusions. Haule, Shim and
Kotliar \cite {Haule08} employed the LDA+DMFT approach
\cite{LDA+DMFT}, the combination of density functional theory (DFT)
in the local density approximation (LDA) with the many-body
dynamical mean-field theory (DMFT) \cite{DMFT}, to study
LaO$_{1-x}$F$_{x}$FeAs. Using a value of the local Coulomb
interaction $U=4$~eV obtained by the random phase approximation
(RPA) \cite{RPA} for metallic iron~\cite {Ferdi} (which is quite
similar to that found by constrained LDA, $U=$ 3--4~eV
\cite{georges,jpcm,physicac}), they concluded that
LaO$_{1-x}$F$_{x}$FeAs is a strongly correlated, bad metal which is
close to a Mott metal-insulator transition. A similar conclusion
was reached by Craco \emph{et al.} \cite {Craco08} for
SmO$_{1-x}$F$_{x}$FeAs. By contrast, using constrained RPA to
compute $U$ in LaFeAsO Nakamura, Arita and Imada \cite {Arita}
obtained smaller values of $U$, in the range $U=$ 2.2--3.3 eV.
Therefore they concluded that in LaFeAsO and LaFePO electronic
correlations are moderately strong. This was affirmed by Anisimov
\emph{et al.} \cite {jpcm} who performed LDA+DMFT investigations for
LaFeAsO, with $U$ taken from constrained LDA \cite{U-calc,anigun},
and compared with photoemission spectroscopy (PES) data. Yet another
conclusion was reached by Yang \emph{et al.} \cite{Yang09} who
investigate the Fe-pnictides by XAS and RIXS and compared with
results from cluster diagonalizations, multiplet and DFT
calculations. They estimated the Coulomb interaction as $U\leq2$ eV
and inferred that the Fe-pnictides are weakly correlated systems.

These different assessments raise a general question about the
classification of the strength of electronic correlations in a
particular material. An overall estimate is provided by the ratio of
the local Coulomb interaction $U$ and the band width $W$. For
$U/W<1$ the system is considered weakly correlated, and the results
of DFT approximations are sufficient to explain its electronic and
magnetic properties. By contrast, if $U$ is comparable with $W$ or
larger, the system is moderately or even strongly correlated, and
the Coulomb interactions must be treated explicitly in electronic
structure calculations. However, the single parameter $U/W$ is only
a rough measure of the electronic correlations; furthermore it is
not directly accessible by experiment. Here the effective mass
renormalization $m^*/m$ of quasiparticles is a more useful parameter
since it permits experimental tests. Nevertheless a single number
cannot reliably assess the correlation strength in a real material
whose properties are determined by various different aspects such as
the electronic band structure, degeneracies, filling, and the
Coulomb interaction parameters beyond $U$. A much more expressive
quantity is the $\bf k$-resolved single-particle spectral function
$A({\bf k}, \omega)$, and its $\bf k$-integrated variant
$A({\omega)}$, both of which can be measured by angular resolved
photoemission spectroscopy (ARPES) and PES, respectively. They allow
for an unambiguous classification of the strength of electronic
correlations. Namely, if $A({\bf k}, \omega)$ is well described by
DFT approximations it is justified to call the system \emph{weakly}
correlated. If the band structure is substantially renormalized by
correlation effects as expressed by an enhancement of the
quasiparticle effective mass, but $A({\omega)}$ does not yet show
incoherent Hubbard bands, the system is \emph{moderately}
correlated. Only if the correlation induced transfer of spectral
weight is so strong that pronounced Hubbard bands and a distinct
quasiparticle peak at the Fermi energy appear, can the system be
called \emph{strongly} correlated (a good example is the 3$d^1$
system SrVO$_3$ \cite{SrVO3}). The spectral functions of correlated
electron materials can be computed within the LDA+DMFT framework
\cite{LDA+DMFT}, which provides access to material specific
single-particle spectra and higher correlation functions.

In this Letter we employ the LDA+DMFT scheme \cite{LDA+DMFT} to explore the
importance of electronic correlation effects in BaFe$_{2}$As$_{2}$,
the undoped parent compound of one of the main pnictide high-$T_c$
superconductor families. For this purpose we calculate the spectral
functions $A({\omega)}$ and $A({\bf k}, \omega)$ and compare our
results with the available experimental data for BaFe$_{2}$As$_{2}$.
In particular, we apply the scheme described above to classify the
correlation strength in this important class of Fe-pnictides.

The LDA+DMFT approach proceeds as follows: First, an effective
Hamiltonian $\Tilde{h}_{\bf k}$ is constructed using converged LDA
results for the system under investigation, then the many-body
Hamiltonian is set up, and finally the corresponding self-consistent
DMFT equations are solved. By projecting onto Wannier functions
\cite{projection} we obtain an effective 16-band Hamiltonian which
incorporates five Fe {\it d} orbitals and three As {\it p} orbitals
for each of the two Fe and As ions per unit cell. In the present
study we construct Wannier states for a single energy window
including both {\it p} and {\it d} bands. Consequently, the
eigenvalues of the effective Hamiltonian $\Tilde{h}_{\bf k}$ exactly
correspond to the 16 Fe and As bands from LDA. Thereby hybridization
effects between As {\it p} and Fe {\it d} electrons are explicitly
taken into account. The LDA calculations were performed with the
experimentally determined crystal structure \cite{Huang} using the
Elk full-potential linearized augmented plane-wave (FP-LAPW) code
\cite{Elk}. Parameters controlling the LAPW basis were kept to their
default values. The calculated LDA band structure
$\epsilon_{LDA}({\bf k})$ is found to be in good agreement with that
of Fink \emph{et al.} \cite{Fink}.

To account for the Coulomb interaction already present in LDA
the $dd$-diagonal elements of the effective Hamiltonian
$\Tilde{h}_{\bf k}$ are renormalized by a double counting correction
\cite{LDA+DMFT}
$E_{dc}=U(n_{\rm dmft}-\frac{1}{2})$,
\begin{equation}
h^{dd}_{\mathbf{k},\alpha\beta}=\Tilde{h}^{dd}_{\mathbf{k},\alpha\beta}-
E_{dc}\delta_{\alpha\beta}.
\end{equation}
Here $n_{\rm dmft}$ is the total, self-consistently calculated
number of {\it d} electrons per Fe site obtained within the
LDA+DMFT. This form of $E_{dc}$ yields reliable results for
transition metal compounds, including the superconductor LaFeAsO
\cite{jpcm}. The $p-d$ Hamiltonian to be solved by DMFT then has the
form
\begin{equation}
\label{eq:ham}
\begin{split}
H=&\sum_{\mathbf{k},\sigma}\bigl(h_{\mathbf{k},\alpha\beta}^{dd}d_{\mathbf{k}\alpha\sigma}^{\dagger}
d_{\mathbf{k}\beta\sigma}+h_{\mathbf{k},\gamma\delta}^{pp}p_{\mathbf{k}\gamma\sigma}^{\dagger}
p_{\mathbf{k}\delta\sigma}+ \\&
h_{\mathbf{k},\alpha\gamma}^{dp}d_{\mathbf{k}\alpha\sigma}^{\dagger}
p_{\mathbf{k}\gamma\sigma}+h_{\mathbf{k},\gamma\alpha}^{pd}p_{\mathbf{k}\gamma\sigma}^{\dagger}
d_{\mathbf{k}\alpha\sigma}\bigr)+\\&\sum_{i,\sigma,\sigma'}U_{\alpha\beta}^{\sigma\sigma'}n^d_{i\alpha\sigma}n^d_{i\beta\sigma'}.
\end{split}
\end{equation}
Here $d_{\mathbf{k}\alpha\sigma}$ and $p_{\mathbf{k}\gamma\sigma}$
are Fourier transforms of $d_{i\alpha\sigma}$ and
$p_{i\gamma\sigma}$, which annihilate the $d$ or $p$ electron with
orbital and spin indices $\alpha\sigma$ or $\gamma\sigma$ in the
$i$th unit cell, and $n^d_{i\alpha\sigma}$ is the corresponding
occupation number operator.

The DMFT self-consistency equations were solved iteratively for
imaginary Matsubara frequencies. The auxiliary impurity problem was
solved by the Hirsch-Fye quantum Monte Carlo (QMC) method
\cite{QMC}. The elements of $U_{\alpha\beta}^{\sigma\sigma'}$ matrix
are parameterized by $U$ and $J$ according to procedure described in
\cite{LichtAnisZaanen}. In the following we use the interaction
parameters $U$=3.1~eV and $J$=0.81~eV obtained in
\cite{jpcm,physicac}. Calculations were performed in the
paramagnetic state at the inverse temperature $\beta=1/T$=20
eV$^{-1}$ . The imaginary time interval $0<\tau <\beta$ was divided
into 150 slices, and 6$\times 10^{6}$ QMC sweeps were used in a
self-consistency loop within the LDA+DMFT scheme. The local
self-energy $\Sigma (\omega)$, which is formally an 16$\times$16
matrix with the only nonzero elements on the diagonal of the {\it
dd} block, was calculated for real energies $\omega$ by analytic
continuation from imaginary Matsubara frequencies  using the Pad\'e
approximant \cite{Pade}.

A quantitative measure of the electron correlation strength is also
provided by the quasiparticle renormalization factor
$Z=(1-\frac{\partial \Sigma}{\partial\omega}|_{\omega=0})^{-1}$
which leads to an effective mass enhancement $m^{*}/m=Z^{-1}$. In
general, the self-energy is a matrix, leading to different effective
masses for different bands. The masses can be obtained from the
self-energy on the real axis, $\Sigma (\omega)$. The calculated
$m^{*}/m$ values for every {\it d}-orbital are presented in Table 1.
The d$_{x^{2}-y^{2}}$ orbital has the smallest effective mass
renormalization ($m^{*}/m$ = 1.83). The other {\it d}-orbitals have
a slightly larger value ($m^{*}/m$ = 2.05 -- 2.07). The overall
effective mass enhancement $m^{*}/m\approx$ 2 agrees well with the
ARPES data for BaFe$_{2}$As$_{2}$ \cite{ding,Yi}. It is also in good
agreement with experimental results for the pnictide material
LaFePO, where de Haas-van Alphen experiments \cite{dHvA} found a
mass enhancement between 1.7 and 2.1, and ARPES \cite{kur5} measured
a band renormalization by a factor of 2.2 compared with the LDA.

\begin{table}
\caption{
Effective mass renormalization $m^{*}/m$ of quasiparticles in
BaFe$_{2}$As$_{2}$ for different orbitals of the {\it d} shell.}
\vspace{3.0mm} \centering
\begin{tabular}{c|cccc}
\hline \hline
Orbitals  & d$_{xy}$ & d$_{yz,xz}$ & d$_{3z^{2}-r^{2}}$ & d$_{x^{2}-y^{2}}$ \\
\hline
$m^{*}/m$  & 2.06     & 2.07     & 2.05               & 1.83              \\
\hline \hline
\end{tabular}
\end{table}

\begin{figure}
\centering \vspace{0.0mm}
\includegraphics[width=0.9\linewidth]{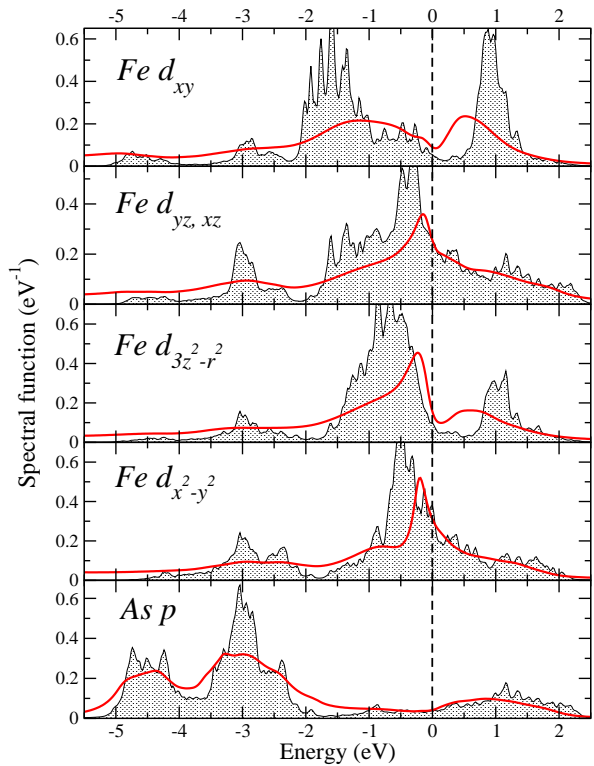}
\caption{(Color online) Orbitally resolved Fe 3{\it d} and total
As {\it p} spectral functions of BaFe$_{2}$As$_{2}$ obtained within
LDA+DMFT (solid lines) are compared with LDA results (shaded areas).}
\label{DMFTvsLDA}
\end{figure}

The orbitally resolved Fe 3{\it d}  and total As 4{\it p} spectral
functions computed within the LDA and LDA+DMFT, respectively, are compared in Fig.
\ref{DMFTvsLDA}. The tetragonal crystal field splitting of the
Fe-$3d$ states in this material is rather weak
($\Delta_{cf}$=0.25~eV). Within the LDA all five Fe $d$ orbitals
form a common band in the energy range ($-$2, $+$2)~eV relative to
the Fermi level (band width $W\approx$  4~eV). There is a
significant hybridization of the Fe $t_{2g}$ orbitals with the As
$p$ orbitals, leading to spectral weight in the energy interval
($-$3, $-$2)~eV where the As $p$ band is located. However, this is
not a correlation effect. The correlations only lead to some
broadening but not to substantial spectral weight transfer, i.e.,
there are no Hubbard bands found in the entire spectrum. Indeed, the
overall shape of the LDA spectral function is hardly changed by the
correlations.

\begin{figure}
\centering \vspace{5.0mm}
\includegraphics[width=0.9\linewidth]{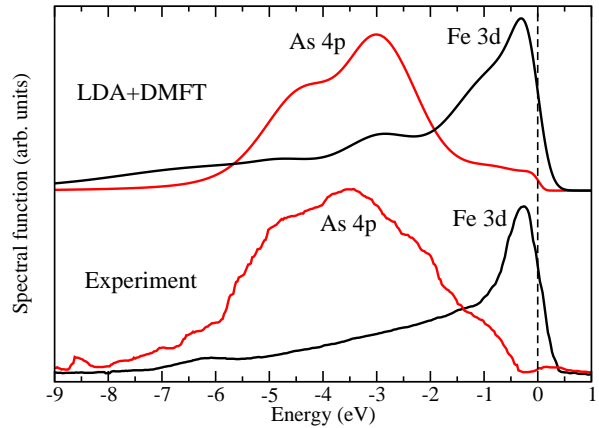}
\caption{(Color online) Normalized total Fe {\it d} and As {\it p}
spectral functions of BaFe$_{2}$As$_{2}$ calculated within LDA+DMFT
(upper curves) are compared with photoemission spectroscopy data of
de Jong {\it et al}. \cite{deJong} (lower curves).}
\label{DMFT_DOS}
\end{figure}

In Fig.\ref{DMFT_DOS} the total Fe 3{\it d} and As 4{\it p} spectral
functions are compared with photoemission spectroscopy data of de
Jong {\it et al}. \cite{deJong}. The computed spectra reproduce the
experimental features very well, including the relative positions of
the {\it d} and {\it p} bands and the strong {\it d} contribution
at the top of the valence band.

\begin{figure}
\centering \vspace{0.0mm}
\includegraphics[width=0.9\linewidth]{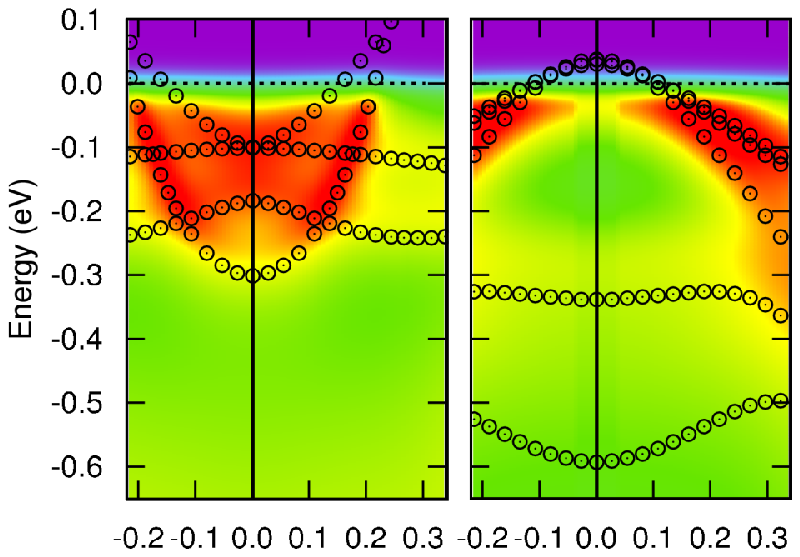}
\includegraphics[width=0.9\linewidth]{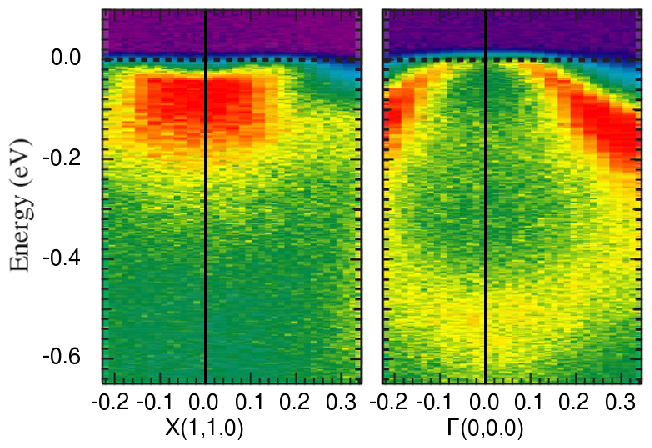}
\caption{(Color online) The {\bf k}-resolved total spectral function
A({\bf k}, $\omega$) of BaFe$_{2}$As$_{2}$
near the $\Gamma$ and $X$ points in the Brillouin zone is depicted
as a contour plot. Upper panel: LDA+DMFT spectral function including
the renormalized band structure (circles) obtained by plotting the
peak positions of the spectral function $A({\bf k}, \omega)$.
Lower panel: The corresponding experimental ARPES intensity map of
Liu {\it et al}. \cite{Liu}.} \label{Ake_contour}
\end{figure}

We now calculate the {\bf k}-resolved spectral function
\begin{equation}
A({\bf k}, \omega)=-Im\frac{1}{\pi} Tr[(\omega+\mu)\hat{I}-\hat h_{\bf k}-\hat{\Sigma} (\omega)]^{-1}.
\label{SFunction}
\end{equation}
Here $\hat h_{\bf k}$ is the 16$\times$16 Hamiltonian matrix on a
mesh of {\bf k}-points and $\mu$ is the self-consistently determined
chemical potential. In Fig. \ref{Ake_contour} we compare our results
with ARPES data of Liu {\it et al}. \cite{Liu}. Both theory and
experiment show dispersive bands crossing the Fermi level near the
$\Gamma$ and $X$ points. In addition, two weakly pronounced
relatively flat bands can be seen in the region from -0.2 to -0.3 eV
and from -0.5 to -0.6 eV near the $\Gamma$ point. The calculated
shape and size of the electron and hole pockets centered at the X
and $\Gamma$ points, respectively, are in good agreement with the
ARPES data. Also shown in Fig. \ref{Ake_contour} (upper panel) is
the correlated band structure $\epsilon_{DMFT}({\bf k})$. Near the
Fermi energy, i.e., in the energy range from -0.25 eV to zero where
quasiparticles are well defined (as expressed by a linear behavior
of Re$ \Sigma(\omega)$), this dispersion is very well represented by
the scaling relation $\epsilon_{DMFT}({\bf k})=\epsilon_{LDA}({\bf
k})/(m^*/m)$, with $m^*/m$ as the computed mass enhancement. This
shows that the band structure is substantially renormalized.
Nevertheless, since there is no substantial spectral weight transfer
in the spectral function, electronic correlations in
BaFe$_{2}$As$_{2}$ are only moderately strong.

In conclusion, by employing the many-body dynamical mean-field
theory in combination with LDA electronic band structure
calculations (LDA+DMFT approach) we calculated the single-particle
{\bf k}-integrated and {\bf k}-resolved spectral functions for
BaFe$_{2}$As$_{2}$, a parent compound of the superconducting iron
pnictide family. The correlated band structure obtained within
LDA+DMFT is well described by the relation $\epsilon_{DMFT}({\bf
k})=\epsilon_{LDA}({\bf k})/(m^*/m)$, with an effective mass
enhancement computed as $m^{*}/m\approx$ 2. At the same time
correlations do not lead to substantial transfer of spectral weight
in the spectrum, i.e., Hubbard bands do not form. Our results are in
good agreement with PES experiments of de Jong {\it et al.}
\cite{deJong} and ARPES data of Liu {\it et al.} \cite{Liu}.  We
also discussed a general classification scheme for the electronic
correlation strength based on the renormalization of the band
structure and the transfer of spectral weight. Our results obtained
for BaFe$_{2}$As$_{2}$ and their comparison with experimental data
show that this material, and also other Fe-pnictides, should be
classified as \emph{moderately correlated}.

The authors thank Jan Kune\v{s} for providing his DMFT(QMC) computer
code used in our calculations. Support by the Russian Foundation for
Basic Research under Grant No. RFFI-07-02-00041, the Dynasty
Foundation, the fund of the President of the Russian Federation for
the support of scientific schools NSH 1941.2008.2, the Program of
the Russian Academy of Science Presidium ``Quantum microphysics of
condensed matter'' N7, and the Deutsche Forschungsgemeinschaft
through SFB 484 is gratefully acknowledged.

\end{document}